\documentclass[12pt, letterpaper]{article}
\usepackage[utf8]{inputenc}

\usepackage{mathptmx}       
\usepackage{helvet}         
\usepackage{courier}        
\usepackage{type1cm}        
\usepackage{makeidx}         
\usepackage{multicol}        
\usepackage[bottom]{footmisc}

\usepackage{amsfonts}
\usepackage{amsmath}
\usepackage{amssymb}
\usepackage{graphicx}

\usepackage{color}
\usepackage{url}

\def\bor{\mathop{\mathord{\lor}\!\!\!\raise4pt\hbox{$\scriptscriptstyle 2$}\,}}
\def\band{\mathop{\mathord{\land}\!\!\!\lower2pt\hbox{$\scriptscriptstyle 2$}\,}}

\def\lambdabar{\protect\@lambdabar}
\def\@lambdabar{%
\relax
\bgroup
\def\@tempa{\hbox{\raise.73\ht0
\hbox to0pt{\kern.25\wd0\vrule width.5\wd0
height.1pt depth.1pt\hss}\box0}}%
\mathchoice{\setbox0\hbox{$\displaystyle\lambda$}\@tempa}%
{\setbox0\hbox{$\textstyle\lambda$}\@tempa}%
{\setbox0\hbox{$\scriptstyle\lambda$}\@tempa}%
{\setbox0\hbox{$\scriptscriptstyle\lambda$}\@tempa}%
\egroup
}

\newtheorem{postulate}{Postulate}
\newtheorem{definition}{Definition}

\newcommand{\CHAIN}[1]{\mathbf{#1}}
\renewcommand{\P}{{\CHAIN{P}}}
\newcommand{\Q}{{\CHAIN{Q}}}

\newcommand{\llseq}[4]{\ensuremath{[#1,#2,#3,#4]}}   

\makeindex             

\title{Understanding the Electron}

\author{Kevin H. Knuth \\Departments of Physics and Informatics\\University at Albany, Albany NY USA}
\date{October 2, 2015}

\begin{document}
%
%
\maketitle

\abstract{Well over a century after the discovery of the electron, we are still faced with serious conceptual issues regarding precisely what an electron is.  Since the development of particle physics and the Standard Model, we have accumulated a great deal of knowledge about the relationships among various subatomic particles. However, this knowledge has not significantly aided in our understanding of the fundamental nature of any particular elementary subatomic particle.  The fact that many particle properties, such as position, time, speed, energy, momentum, and component of spin, are observer-dependent suggests that these relevant variables do not represent properties per se, but rather the relationship between the observer and the observed.  That is, they reflect details about how the electron influences the observer, and vice versa.  Here we attempt to understand this by considering a simple model where particles influence one another in a discrete and direct fashion.  The resulting framework, referred to as Influence Theory, is shown to faithfully reproduce a surprising amount of physics.  While it would be naive to assume that the ideas presented here comprise anything resembling the final word on the matter, it is hoped that this work will demonstrate that a simple and understandable
picture of particles, such as the electron, is indeed feasible and should be actively sought after.}

\section{Introduction}
\label{sec:1:Introduction}
Whether it is the crack and snap of an electric shock on a cold winter day or the boom and crash of a lightning bolt on a stormy summer afternoon, we are familiar with electrons because they influence us.  Similarly, scientists know about electrons because they influence their measurement equipment.  Electrons are described in terms of properties inspired by our descriptions of billiard balls, such as mass, position, energy, and momentum, as well as additional non-billiard-ball-like properties such as charge and spin.  The laws of physics, as applied to electrons, focus on describing the interrelationships among these relevant variables as well as their relationships to external forces.

Since the discovery of the electron \cite{Thomson:1897:cathode-rays}, we have become so familiar with these electron properties that we have come to view them as foundational.  This is despite the fact that there are well-known serious conceptual problems that have plagued even the most prominent physicists over the last century.  For example, Einstein, when faced with the burgeoning field of particle physics, was compelled to write, ``You know, it would be sufficient to really understand the electron.''  This is a difficult issue since quantum mechanics tells us that in some situations an electron acts as a particle and in others it acts as a wave.  Moreover, quantum mechanics also tells us that an electron cannot simultaneously possess both a definite position and a definite momentum.  While this may not be so surprising given the familiarity that today's physicists have with quantum mechanics and the fact that the vast majority have simply come to accept that this is how things are, the conceptual issues remain unsolved.  Now it is perhaps less well known that Breit \cite{Breit:1928} and Schr{\"o}dinger \cite{Schrodinger1930:kraftefreie} both independently showed that the Dirac equation, which most accurately describes the behavior of a single electron, results in velocity eigenvalues of $v = \pm c$.  That is, at the finest of scales, the electron can \emph{only} be observed to move the speed of light! Schr{\"o}dinger called this phenomenon \emph{Zitterbewegung}, or shuddering motion, since the implication is that an electron must either zig-zag back-and-forth or spiral at the speed of light---even when at rest \cite{Huang:1952zitterbewegung}\cite{Hestenes:1990zitterbewegung}\cite{Hestenes:1993}\cite{Hestenes:2008electron}\cite{Hestenes:2010zitterbewegung}.  Not only does this bring into question what is meant by the state of rest, but it is also seemingly contrary to what we expect from a massive particle within the context of special relativity. And yet, the Dirac equation describes the relativistic quantum mechanics of a single electron. However, we do not need to go so far as quantum mechanics to identify lingering conceptual problems.  For instance, most of the electron properties are observer-dependent: position, time, speed, energy, momentum, component of spin, which strongly suggests that they are not \emph{properties possessed} by an electron, but rather they are somehow a \emph{description of the relationship} between the electron and the observer.

Physicists studying foundations must carefully consider the implications of such clues and conceive of theoretical models that accommodate them.  This is not accomplished by assuming high-level concepts such as Lorentz invariance, Hilbert spaces, and Lagrangians, since there exists no foundational rationale for such detailed technical assumptions---save for the current conceptual formulation of physics, which we aim to understand more deeply.  Likewise, principles involving the relationships among relevant variables, such as energy and momentum in the Principle of Conservation of Four-Momentum or position and momentum in the Principle of Complementarity, will not serve us either since the adoption of such principles precludes us from going deeper and understanding the nature of the relevant variables themselves. One cannot understand why something is true by assuming it to be true.  Instead, we must ask the question: What is the nature of the electron properties such that they behave the way they do, and is there perhaps a simpler way to think about them?

In this paper, I present a simple model of the electron and show that it faithfully reproduces a surprising amount of physics.  It would be naive to assume that the ideas presented here comprise the final word on the matter.  Instead, it is hoped that this work will demonstrate that a simple and understandable picture of elementary particles, as well as a foundation for physics, is indeed feasible and should be actively sought after.

\section{Electron Properties}
\label{sec:2:Properties}
Let us begin by performing a thought experiment.  Imagine that electrons are pink and fuzzy, but that their pinkness and fuzziness do not affect how they influence any measurement apparatus.  Since these hypothetical electron properties are undetectable by measurement, they are forever inaccessible to us as experimenters.  That is, there is no way for us to know about the pink and fuzzy nature of the electron.  Turning this thought on its side, it becomes readily apparent that the only properties of an electron that we as experimenters have access to are those that affect how an electron influences our equipment.  In fact, an operationalist definition of such properties would consist simply of a description of the effects of the electron's influence.

This is, in fact, how one builds up the development of the theory of magnetism in a class lecture.  The magnetic force is defined by how charges act in a variety of situations.  We now know, since relativity, that the magnetic force itself is observer-dependent in that in a different reference frame it can be perceived as a combination of an electric and magnetic force, or in the limiting case, simply an electric force.  For this reason, the modern perspective involves the concept of the electromagnetic force where the electric or magnetic nature of the force is dependent on the relationship between the system and the observer.

The observation that many electron properties are observer-dependent suggests that what we think of as properties of the electron more accurately reflect the relationship between the electron and the observer.  However, it is not clear how many of these properties might be fundamentally unified and yet differentiate themselves in certain situations only because of a particular relation to an observer.  Moreover, what is this electron-observer relation, and how does it give rise to the physics that we are familiar with?  Whatever the result of such inquiry, we can be certain that we will have to abandon one or more firmly held concepts related to ``particle properties'' in favor of something more fundamental.

\section{Influence}
\label{sec:3:Influence}
Perhaps instead of focusing on what an electron \emph{is} and what properties an electron \emph{possesses}, it would be better to simply focus on what an electron \emph{does} to an observer to define its \emph{relationship} to that observer.  Since we know for certain that electrons can influence our equipment and that our equipment can influence electrons, we focus on this simple fact and introduce the concept of \emph{influence}, referring to the resulting theoretical framework as \emph{Influence Theory}.

Inspired in part by Wheeler and Feynman \cite{Wheeler+Feynman:1945}\cite{Wheeler+Feynman:1949}, we begin with the simple assumption: \emph{an elementary particle, such as an electron, can influence another particle in a direct and discrete fashion.}  One might be tempted to refer to this as a direct particle-particle interaction.  However, the term interaction implies something bi-directional, whereas influence here is assumed to be directional---something directed from source to target.  As such, each \emph{instance of influence} enables one to define two events: \emph{the act of influencing} and \emph{the act of being influenced}.  The act of influencing is associated with the source particle, and the act of being influenced is associated with the target particle.

We imagine the observer to possess an instrument that can monitor an elementary particle and provide information about that particle to the observer.  When we say that the observer was influenced by the electron, what we will mean is that the monitored particle was influenced by the electron and that the observer detects the fact that the monitored particle was influenced.  Since the aim is to describe the electron, we will take its perspective and refer to the act of influencing associated with the electron as an \emph{outgoing or emitted influence event}. Similarly, the monitored particle may influence the electron.  This also is assumed to be detectable, and again, taking the perspective of the electron, we will refer to this act of being influenced associated with the electron as an \emph{incoming or received influence event}.

At this point, the reader is most likely questioning the nature of such influence.  We should immediately dispel any notion of propagation and state emphatically that we do not assume that influence propagates through space and time from source to target.  Instead, influence simply relates one particle to another in the sense that one particle influenced and the other was influenced.  By assuming that only the occurrence of influence is detectable, any properties an instance of influence may possess remain inaccessible.  What is remarkable is that we will find that the potentially inaccessible nature of influence does not matter, since what one may erroneously think of as properties of an electron will be shown to emerge as unique consistent descriptions of the influence-based relationship between the electron and the observer. As a result, we will demonstrate that this proposed concept of influence is potentially responsible for the traditional concepts of position, time, motion, energy, momentum, and mass, as well as several well-known quantum effects.

Here we summarize the postulates that provide the foundation for the model.
\begin{postulate}
Elementary particles can influence one another in a pairwise, directed, and discrete fashion, such that given an instance of influence, one particle influences, and the other is influenced.
\end{postulate}
This postulate allows us to define the concept of an event.
\begin{definition}[Event]
Every instance of influence results in two \emph{events}, each associated with a different particle: \emph{the act of influencing} and \emph{the act of being influenced}.
\end{definition}
By virtue of the fact that these two events can be distinguished, they can be ordered.  Consistently choosing the way in which we order the two influence events defines a binary ordering relation $<_i$ that acts on pairs of events, each pair defined by a single influence instance. The subscript $i$ indicates that this ordering is due to an influence instance.

The following postulates together enable one to order the events associated with a single particle.  To accomplish this, we assume that the particle has some internal, potentially inaccessible, state that is somehow coupled to the influence instances.  We keep this minimal by assuming no details about their relationship. We simply assume that a relationship exists.\footnote{If all one can detect is the occurrence of an influence event, how can anything ever be known about the relation between those events and any internal states of the particle?}
\begin{postulate}
A particle has associated with it a potentially inaccessible internal state such that each influence event couples one particle state to one other particle state.  It is in this sense that each influence event is bounded by two particle states.
\end{postulate}
We also assume that the influence instances are coupled to particle states.
\begin{postulate}
Each particle state that bounds an influence event couples that influence event to one another influence event associated with the same particle, such that each particle state is bounded by two influence events.
\end{postulate}
These two postulates allow one to totally order the influence events associated with a particle, as well as the particle states, with a transitive binary ordering relation, which we shall denote as $<_c$ where the subscript $c$ denotes coupling through internal particle states.  This results in one being able to describe a particle in terms of a totally ordered chain of events.\footnote{If the particle states were accessible, then we could alternatively describe the particle as a totally ordered chain of particle states.}

A further consequence of these postulates is the fact that each event along a particle chain is either the result of that particle influencing another or of that particle being influenced.  We can take this further and define a new ordering relation $<$ based on considering the two ordering relations $<_i$ and $<_c$ to belong to an equivalence class.
\begin{definition}[Generic Ordering]
Two events $x$ and $y$ are \emph{ordered} with $x$ included by $y$, denoted by $x < y$, if $x <_i y$ or $x <_c y$, or some transitive combination of $<_i$ and $<_c$, such that the arbitrary directions of the binary ordering relations $<_i$ and $<_c$ are selected to avoid cyclic relationships.  We can more generally write $x \leq y$ where either $x < y$ or $x = y$.
\end{definition}

Together, these postulates result in a model of particle behavior summarized by a set of events, which can be compared using a transitive binary ordering relation defined by the process of influence.  This results in an acyclic graph, or a partially-ordered set (poset for short).  If one conceives of the ordering as being the foundation of causality, then this is analogous to a causal set \cite{Bombelli-etal-causal-set:1987}\cite{Sorkin:2003}\cite{Sorkin:2006} where the events are causally ordered, but with a specific connectivity.  In this framework, a given particle is described by an ordered sequence, or chain, of events.  Each event on one particle chain either covers\footnote{An event $z$ covers an event $x$ if $x < z$ and there does not exist any $y$ such that $x < y$ and $y < z$.} or is covered by precisely one event on another particle chain.  We do not assume that these events take place in any kind of space or time---only that they can be partially ordered.

Given this purposefully simplistic model of a particle, such as an electron, we proceed by developing all aspects of the theory from the bottom up.  This work combines the results of several previous efforts that rely on the consistent quantification of systems based on algebraic symmetries \cite{Knuth:laws}\cite{Knuth:FQXI2015}\cite{Knuth&Skilling:2012}\cite{GKS:PRA}\cite{GK:Symmetry}\cite{Knuth+Bahreyni:JMP2014}\cite{Knuth:Info-Based:2014}\cite{Knuth:FQXI2013}\cite{Knuth:MaxEnt2014:motion}\cite{Walsh+Knuth:acceleration}, and we will refer the reader to these works for more details.  The idea is to work through each step in sufficient detail to illustrate how quantification of order-theoretic structures enables one to derive laws that reproduce a surprising amount of physics.  That is, rather than postulating laws and perceiving them as representing some kind of underlying natural order, we instead postulate the nature of the underlying order and derive the resulting laws.

\section{Coarse-Grained Picture of Influence}

\subsection{Intervals: Duration and Directed Distance}
\label{sec4:Position-Time}
It has been posited that an observer monitoring a particle detects events in an ordered sequence, which one can think of as pre-time (ordering without scale).  Rather than focusing on the precise poset of events generated by a set of particles influencing one another, we begin by considering a coarse-grained picture of a poset.  This is accomplished by defining an order-preserving map that takes a set of successive events along a particle chain to a single element.\footnote{As an example, given events $p_1 < p_2 < p_3 < \ldots < p_{12}$ along the chain $\P$, the map $\phi$ which gives $\phi(p_1) = \phi(p_2) = \phi(p_3) < \phi(p_4) = \phi(p_5) = \phi(p_6) < \phi(p_7) = \phi(p_8) = \phi(p_9) < \phi(p_{10}) = \phi(p_{11}) = \phi(p_{12})$ is a valid coarse-graining map.}  This will result in a poset of coarse-grained events, each of which consists of multiple fundamental events along with multiple influences.  The point of this will be to demonstrate that there exists a unique means (up to scale) by which an embedded observer represented by a chain of events in the poset can quantify a subset of poset and that this is equivalent to the mathematics of space-time.

\subsubsection{Quantification of a Chain}
\begin{figure}[t]
  \begin{center}
  \includegraphics[height=0.30\textheight]{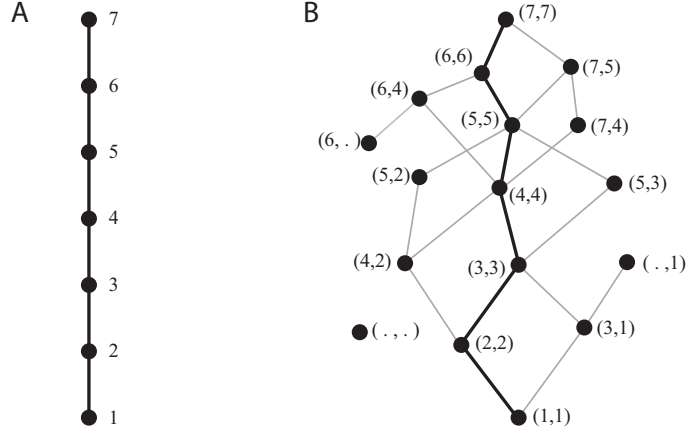}
  \end{center}
  \caption{A. A quantification of a finite chain is performed by assigning a monotonic valuation to the events.  In this case, we simply count the events using natural numbers.  B. Any event from the set of accessible events can be quantified with respect to an observer by either one or two numbers by forward projecting and backward projecting onto the observer chain (depending on whether those projections exist). }
  \label{fig:fig-1}
\end{figure}

We begin by introducing the concept of quantification, which will allow us to quantify events and their relationships to one another using numbers.  In general, a consistent quantification of the events comprising the observer chain (either fine-grained or coarse-grained) is given by a monotonic valuation, which is a real-valued function $q$ that acts as an order-isomorphism taking each event to a real number such that if the event $y$ includes the event $x$, $x \leq y$, then the real-valued quantities $q(x)$ and $q(y)$ assigned to events $x$ and $y$, respectively, are related by $q(x) \leq q(y)$ (see Figure \ref{fig:fig-1}A).  The idea is that the quantification captures the ordering of the events experienced by the observer.

\subsubsection{Incomparability and Inaccessibility}
If it is true that all elements of a chain $\P$ are incomparable to an event $x$ in the poset\footnote{The event $x$ is said to be incomparable to the event $y$ if it is true that $x \nleq y$ and $y \nleq x$.}, then we say that the event $x$ is \emph{inaccessible} to the chain $\P$.  On the other hand, if there exists at least one element of the chain that is comparable to the event $x$, then we say that $x$ is \emph{accessible} to the chain $\P$.  This implies that every observer chain divides the poset into two subsets: a set of events that are accessible to the observer and a set of events that are inaccessible to the observer.  Thus the universe of events is naturally divided into the observable universe and the unobservable universe.

\subsubsection{Chain Projection and the Quantification of Accessible Events}
We can extend the concept of quantification of an observer chain to the set of accessible events by means of assigning up to two unique numbers representing the relationship between an accessible event $x$ and the observer chain $\P$.

If the event $x$ is included by an event on the observer chain, that is, if there exists some element $p \in \P$ such that $x \leq p$, then we can define a mapping $P$ called the \emph{forward projection} that takes $x$ to the least element of $\P$ that includes it, which is given by $Px = \inf\{p \in P \; | \; x \leq p\}$.  This allows us to quantify the event $x$ by assigning to it the quantity assigned to the element $Px$ on the chain.  Similarly, if the event $x$ includes an event on the observer chain, such that there exists some element $p \in \P$ such that $p \leq x$, then we can define a mapping $\bar{P}$ called the \emph{backward projection} that takes $x$ to the greatest element of $\P$ that it includes, which is given by $\bar{P}x = \sup\{p \in P \; | \; p \leq x\}$.  This provides a second possible means by which one can quantify the event $x$ by assigning it the quantity assigned to the element $\bar{P}x$ on the chain.  This means that any event accessible to the observer $\P$ can be uniquely quantified by either one or two numbers resulting in a rather strange observer-based coordinate system (Figure \ref{fig:fig-1}B).

\begin{figure}[t]
  \begin{center}
  \includegraphics[height=0.30\textheight]{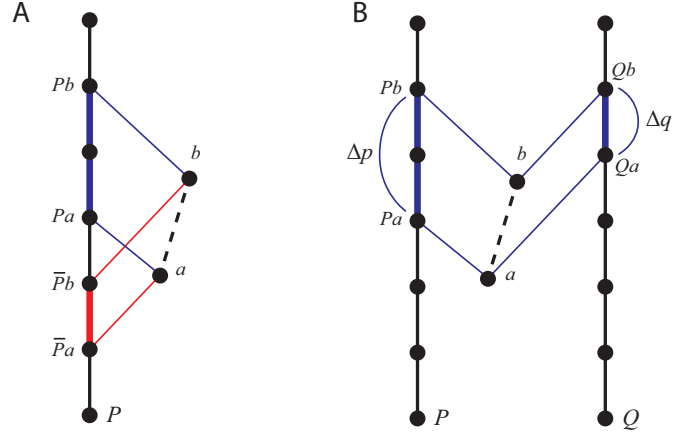}
  \end{center}
  \caption{A. Intervals, which describe the relationship between two events, can be uniquely quantified by chain projection as well. In this example, the interval can be quantified by four numbers, $Pa, Pb, \bar{P}a, \bar{P}b$, determined by the projections of the events defining the endpoints of the interval onto the observer chain; a pair of numbers, $(Pb-Pa,\bar{P}b-\bar{P}a)$, which reflect the lengths of the projections of the interval onto the observer chain; and a scalar, $(Pb-Pa)(\bar{P}b-\bar{P}a)$, given by the product of those lengths. B. Two coordinated observers can be used to quantify an interval using forward projections only.  This also allows one to define a discrete 1+1-dimensional subspace in the poset.  Here the interval is quantified by four numbers, $Pa, Pb, Qa, Qb$; a pair of numbers $(Pb-Pa,Qb-Qa)=(\Delta p, \Delta q)$; and a scalar $(Pb-Pa)(Qb-Qa)=\Delta p \Delta q$, such that the pair and scalar quantifications agree with those obtained using both forward and backward projections onto a single chain as in A.}
  \label{fig:fig-2}
\end{figure}

\subsubsection{Quantification of Intervals}
We can extend this concept of consistent quantification to intervals, which are defined by a pair of events $a$ and $b$ (either comparable or incomparable) and denoted $[a,b]$.\footnote{The two events defining the interval are assumed to be collinear to the coordinated pair of observers. This is precisely defined in \cite{Knuth+Bahreyni:JMP2014} in terms of projections.}  If we restrict ourselves to the special case where both the forward projection and backward projection of both $a$ and $b$ onto the observer chain $\P$ exist, then the forward and backward projections of the pair of events $a$ and $b$ take the interval $[a,b]$ to intervals $[Pa, Pb]$ and $[\bar{P}a, \bar{P}b]$, respectively, on the observer chain (Figure \ref{fig:fig-2}A).  One can prove \cite{Knuth+Bahreyni:JMP2014} that this results in three unique consistent (with respect to rescaling) ways to quantify an interval with respect to an observer:
\begin{align}
&\mbox{Quadruple} &(Pa, Pb, \bar{P}a, \bar{P}b) \\
&\mbox{Pair} &(Pb-Pa, \bar{P}b-\bar{P}a) \\
&\mbox{Scalar} &(Pb-Pa)(\bar{P}b-\bar{P}a)
\end{align}
where the pair is comprised of the lengths of the projected intervals on the observer chain, and the scalar is the product of those lengths.

One can reasonably assume that an observer can only obtain information about an electron if it is influenced by that electron.  This suggests that one can only quantify observer-accessible events by using forward projection.  This gives rise to the concept of a coordinated pair of observers, which is a pair of observers that influence one another in consistent fashion such that the two observers agree on the lengths of intervals on each other's chains.  This turns out to be equivalent to defining a 1+1-dimensional inertial frame \cite{Knuth+Bahreyni:JMP2014}.  Quantification of an interval $[a,b]$ by a pair of coordinated observers $\P$ and $\Q$ using forward projections is illustrated in Figure \ref{fig:fig-2}B.  Again, the interval can be consistently quantified in three ways: \footnote{Please see \cite{Knuth+Bahreyni:JMP2014} for technical details.}
\begin{align}
&\mbox{Quadruple} &(Pa, Pb, Qa, Qb) \\
&\mbox{Interval Pair} &(\Delta p, \Delta q) \\
&\mbox{Interval Scalar} &\Delta p \, \Delta q \label{eq:interval-scalar}
\end{align}
where $\Delta p = Pb-Pa$ and $\Delta q = Qb-Qa$ are the projected lengths of the intervals on the observer chains.

One can transform to a more convenient set of coordinates by considering symmetric and antisymmetric combinations of projected interval lengths:
\begin{align}
\Delta t &= \frac{\Delta p + \Delta q}{2} \label{eq:duration}\\
\Delta x &= \frac{\Delta p - \Delta q}{2}.\label{eq:directed_distance}
\end{align}
The quantity $\Delta t$, which is referred to as \emph{duration}, quantifies intervals that lie along the observer chains (an ordered relationship), and the quantity $\Delta x$, which can be referred to as a \emph{directed distance}, quantifies the relationships between chains  (an unorderable relationship)\footnote{Directed distance differs from distance by at most a sign, which indicates the orientation of the interval with respect to the observers $\P$ and $\Q$.}.  These quantities, $\Delta t$ and $\Delta x$, are related to the interval scalar (\ref{eq:interval-scalar}) by
\begin{align}
\Delta s^2 &= \Delta p \, \Delta q \\
&= \left(\frac{\Delta p + \Delta q}{2} \right)^2 - \left(\frac{\Delta p - \Delta q}{2} \right)^2 \\
&\equiv \Delta t^2 - \Delta x^2,
\end{align}
which is the Minkowski metric in $1+1$ dimensions.  Thus the mathematics of space and time appears to emerge as the unique means by which embedded observers can quantitatively describe the events accessible to them.  We refer to this observer-based description of the poset as the \emph{space-time picture}.

\subsection{Motion and Velocity}
\label{sec5:Motion}
In the previous section, we demonstrated that concepts such as duration and directed distance reflect the relationship between the events and the observers.  Consequently, and perhaps not unexpectedly, these quantities are expected to change when one transforms from one pair of observers to another.

One can consider transforming from one pair of coordinated observers $\P\Q$ to a second pair of coordinated observers $\P'\Q'$ in the more general case where intervals of length $\kappa$ along the chains $\P$ and $\Q$ project to intervals of length $m$ on $\P'$ and intervals of length $n$ on $\Q'$. For the observers to consistently quantify intervals with the interval scalar, we have that
\begin{equation}
\kappa^2 = mn,
\end{equation}
which implies that the interval pair transforms as \cite{Knuth+Bahreyni:JMP2014}
\begin{equation} \label{eq:interval-pair-transform}
(\Delta p', \Delta q') = \left(\sqrt{\frac{m}{n}} \Delta p, \sqrt{\frac{n}{m}} \Delta q\right).
\end{equation}
This implies that the quantity $\Delta t$, quantifying duration along chains, and the quantity $\Delta x$, quantifying directed distance between chains, will mix when transforming to the primed coordinate system.  It is a matter of straightforward algebra to show that the transformation
\begin{align}
\Delta t' &= \frac{\sqrt{\frac{m}{n}}+\sqrt{\frac{n}{m}}}{2} \Delta t + \frac{\sqrt{\frac{m}{n}}-\sqrt{\frac{n}{m}}}{2} \Delta x\\
\Delta x' &= \frac{\sqrt{\frac{m}{n}}-\sqrt{\frac{n}{m}}}{2} \Delta t + \frac{\sqrt{\frac{m}{n}}+\sqrt{\frac{n}{m}}}{2} \Delta x
\end{align}
results in the Lorentz transformation
\begin{align}
\Delta t' &= \frac{1}{\sqrt{1-\beta^2}} \Delta t + \frac{\beta}{\sqrt{1-\beta^2}} \Delta x\\
\Delta x' &= \frac{\beta}{\sqrt{1-\beta^2}} \Delta t + \frac{1}{\sqrt{1-\beta^2}} \Delta x
\end{align}
consistent with special relativity where
\begin{equation}
\beta = \frac{m-n}{m+n}.
\end{equation}

The quantity $\beta$ is immediately recognized as the velocity of the unprimed frame $\P\Q$ with respect to the primed frame $\P'\Q'$.  This is because $m$ and $n$ are the projected lengths, $\Delta p'$ and $\Delta q'$, of an interval of length $\kappa$ on chains $\P$ and $\Q$ onto $\P'$ and $\Q'$. The quantity $\frac{m-n}{2}$ is the directed distance of that interval in the primed frame, and $\frac{m+n}{2}$ is the projected duration in the primed frame, so that in general
\begin{equation}
\beta = \frac{\Delta p' - \Delta q'}{\Delta p' + \Delta q'} \equiv \frac{\Delta x'}{\Delta t'}, \label{eq:beta}
\end{equation}
as expected.

One of the strange results of special relativity is the fact that durations and distances are not concrete fixed physical quantities, but rather they are observer-dependent and can change.  In the context of Influence Theory, this is not strange at all, since the quantities of duration, directed distance, and velocity merely reflect the relationship between the observer and the observed, and changing observers would, in general, be expected to result in a change in these quantities. As a consequence, these results would suggest that space and time are not physical things with properties.  Space and time are instead uniquely consistent descriptions of the relationship between the observed and the observer.

\subsection{Rates: Energy, Momentum and Mass}
\label{sec6:Energy-Momentum}
We have been looking at events along the observer chain in terms of intervals, which led to relevant variables that reflect the concepts of duration, directed distance, and velocity.  With any ordered sequence of elements, there is a dual perspective where one describes the sequence in terms of rates.  Intervals and rates are Fourier transforms of one another and as such, the interval scalar, duration, and directed distance each have Fourier duals, which we show are related to mass, energy, and momentum, respectively.

We define the rates at which a particle influences a pair of coordinated observers in terms of a selected number $N$ of outgoing influence events emitted by the particle chain divided by the duration as measured by an observer.
\begin{equation}
r_P = \frac{N}{\Delta p} \qquad r_Q = \frac{N}{\Delta q}.
\end{equation}
As such, these rates are, again, observer-based.
Combining them in a symmetric and antisymmetric fashion \cite{Knuth:FQXI2013}\cite{Knuth:Info-Based:2014} results in the quantities that we will refer to as \emph{energy}
\begin{equation} \label{eq:energy}
E = \frac{r_P + r_Q}{2},
\end{equation}
\emph{momentum}
\begin{equation} \label{eq:momentum}
p = \frac{r_Q - r_P}{2},
\end{equation}
and \emph{mass}
\begin{equation}
m = \sqrt{{r_P}{r_Q}}
\end{equation}
such that the familiar mass-energy-momentum relationship holds:
\begin{equation} \label{eq:mass-energy-momentum}
m^2 = E^2 - p^2.
\end{equation}

If we assume that the particle has no preference for influencing $\P$ or $\Q$, then we can write
\begin{equation}
\langle \Delta p \rangle = \frac{N}{2} k  \qquad  \langle \Delta q \rangle = \frac{N}{2} \frac{1}{k},
\end{equation}
where the factor $k = \sqrt{\frac{m}{n}}$ from (\ref{eq:interval-pair-transform}) reflects the choice of scale $k=1$ in the rest frame and naturally ensures Lorentz invariance.  We can then write the mass squared as
\begin{equation}
m^2 = r_P r_Q = \frac{N^2}{\langle \Delta p \rangle \langle \Delta q \rangle} = 4
\end{equation}
giving us a mass \footnote{This observation was made by James L. Walsh.} of
\begin{equation} \label{eq:mass=2}
m=2.
\end{equation}
Keep in mind that at this point, this is a single particle theory.  We have not yet considered a model where particles can influence at different rates.

It is straightforward to verify that the velocity, as defined in (\ref{eq:beta}), can be written in terms of energy (\ref{eq:energy}) and momentum (\ref{eq:momentum}) as
\begin{equation}
\beta = \frac{p}{E}.
\end{equation}
Furthermore, since $m^2 = {r_P}{r_Q}$ is an invariant, it is also straightforward to verify that these rate-based quantities transform properly under boosts \cite{Knuth:FQXI2013}\cite{Knuth:Info-Based:2014}.

While the mass-energy-momentum relation (\ref{eq:mass-energy-momentum}) appears here in a new foundational context, such a conception of mass, energy and momentum should not be surprising as it is closely related to the concept of the \emph{internal electron clock rate} hypothesized by de Broglie \cite{deBroglie:1924}\cite{Hestenes:2008electron}.  In his 1924 thesis, de Broglie considered Planck's Law, which already considers energy to be a frequency,
\begin{equation}
E = \hbar\omega,
\end{equation}
and Einstein's Law, which relates energy to mass,
\begin{equation}
E = mc^2,
\end{equation}
and reasoned that mass was related to a frequency
\begin{equation}
m = \frac{\hbar\omega}{c^2}.
\end{equation}

There is no mystery here.  Duration and energy, and directed distance (position) and momentum, are \emph{not complementary properties} of an electron, which cannot be measured accurately simultaneously.  They are instead \emph{complementary descriptions of the relationship} between the electron and the observer.  Duration and directed distance are obtained by considering intervals, whereas energy and momentum are obtained by considering rates, which are necessarily long-term averages.  This aspect of quantum complementarity, along with its reliance on the Fourier transform, emerges naturally as a relationship between the means by which one describes sequences of events.

\section{Fine-Grained Picture of Influence}
Up until this point, we have been considering a coarse-grained picture of influence, which has resulted in well-defined concepts of duration, directed distance, energy, momentum, and mass.  We now return to the fine-grained picture, where every event on the particle chain represents either an act of influence directed at another particle or the act of being influenced by another particle.  The aim is to apply the concepts developed in the previous sections to better understand how a particle, such as an electron, should behave at the most fundamental level from the perspective of Influence Theory.

\subsection{Zitterbewegung}
As is usual in physics, we gain insight by focusing on an ideal situation.  Traditionally, a free particle is a particle that is free from the influence of outside forces.  Here we define a \emph{free particle} as a particle that influences others but is itself not influenced.  As before, we focus on the case of 1+1 dimensions as defined by two coordinated observers where the particle is assumed to be collinear to these observers (see \cite{Knuth+Bahreyni:JMP2014} for technical details).

\begin{figure}[t]
  \begin{center}
  \includegraphics[height=0.275\textheight]{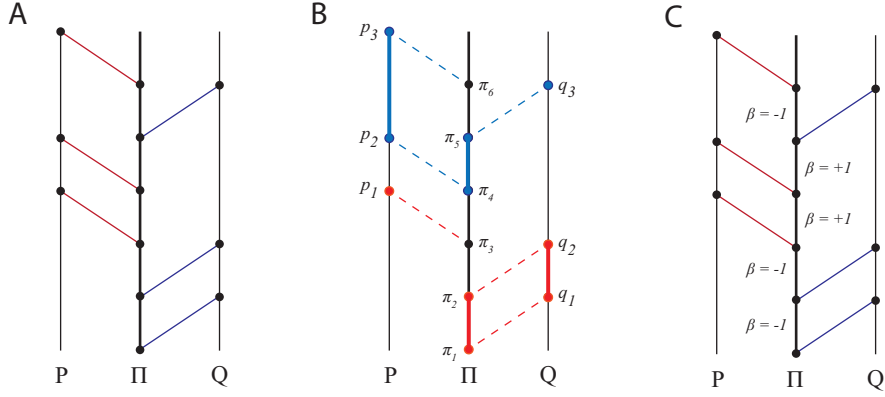}
  \end{center}
  \caption{A. A free particle $\Pi$ that influences the observer chains $\P$ and $\Q$.  Note that the physical distance between events on the sheet of paper is meaningless---it is necessary for the particle chain to be drawn as a straight line.  B. An illustration of the projection of intervals defined by successive events on the chain $\Pi$. The interval $[\pi_1, \pi_2]$ projects onto a degenerate interval $[p_1, p_1]$ of length $\Delta p = 0$ on the chain $\P$ and projects onto the interval $[q_1, q_2]$ of length $\Delta q > 0$ on the chain $\Q$.  Similarly, the interval $[\pi_4, \pi_5]$ projects onto the interval $[p_2, p_3]$, which has length $\Delta p > 0$, and the degenerate interval $[q_3, q_3]$ with length $\Delta q = 0$. C. The lengths of the projections of these intervals along the particle chain onto the observer chains allow one to define an associated velocity.  Here we see that the velocities $\pm 1$ are assigned to each successive interval.}
  \label{fig:fig-3}
\end{figure}

Figure \ref{fig:fig-3}A illustrates a free particle $\Pi$ that influences the chains $\P$ and $\Q$.  Unlike in the coarse-grained case each event on the particle chain is covered by only one event on either the chain $\P$ or the chain $\Q$.  Forward projecting that event to the chain that it influences is trivial.  However, forward projecting to the other chain relies on transitivity through some successive event (if it exists).  This allows one to project intervals defined by successive events on $\Pi$ onto the observers $\P$ and $\Q$ as illustrated in Figure \ref{fig:fig-3}.

One gains significant insight by considering the velocities assigned to the intervals along the particle chain defined by successive events.  Note that any such interval forward projects to one chain resulting in a projected interval of non-zero length, while forward projecting to the other chain resulting in a degenerate interval of zero length (see Figure \ref{fig:fig-3}B).  In the case where the lesser event on the particle chain influences $\P$, we then have that $\Delta p > 0$ and $\Delta q = 0$.  Since the velocity is given by
\begin{align}
\beta &= \frac{\Delta x}{\Delta t} \\
&= \frac{\Delta p - \Delta q}{\Delta p + \Delta q}
\end{align}
we have that $\beta = +1$.  Similarly, in the case where the lesser event on the particle chain influences $\Q$, we have that $\Delta p = 0$ and $\Delta q > 0$ so that $\beta = -1$.  Thus when the particle influences $\P$, the observers describe it as moving to the right at the ultimate speed.  Similarly, when the particle influences $\Q$, the observers describe it as moving to the left at the ultimate speed.

A free particle influencing to the right and left with non-zero rates, $r_P$ and $r_Q$, has a mass $m = \sqrt{r_P r_Q}$.  In this case, every observer describes the particle as zig-zagging back-and-forth at the speed of light, with probabilities of moving left and right given by
\begin{align}
Prob(R) \equiv Prob(\beta = +1) &= \frac{\Delta p}{\Delta p + \Delta q}\\
&= \frac{r_Q}{r_P + r_Q}
\end{align}
and
\begin{align}
Prob(L) \equiv Prob(\beta = -1) &= \frac{\Delta q}{\Delta p + \Delta q}\\
&= \frac{r_P}{r_P + r_Q}.
\end{align}
One can find the average velocity, $\langle \beta \rangle$, by
\begin{equation} \label{eq:beta_in_terms_of_probs}
\langle \beta \rangle = Prob(R) - Prob(L),
\end{equation}
or equivalently one can write the probabilities in terms of the average velocity as
\begin{equation}
Prob(R) = \frac{1+\langle \beta \rangle}{2}
\end{equation}
and
\begin{equation}
Prob(L) = \frac{1-\langle \beta \rangle}{2}.
\end{equation}
Note that the rates $r_P$ and $r_Q$ transform inversely under boosts, so that the mass is invariant.  However, this changes the probabilities with which the particle is observed by a given observer to move left or right, which results in a change of average velocity $\langle \beta \rangle$, as well as energy and momentum, as expected when one keeps in mind that these are all observer-based quantities reflecting a description of the relationship between the particle and observer.  Despite the fact that it has been demonstrated that this behavior is consistent with special relativity on average \cite{Knuth:MaxEnt2014:motion}, it is curious that not even special relativity can describe what the universe looks like to these particles at the finest of scales, since relativity does not describe what happens when a particle moves at the speed of light.

It is interesting to consider the limiting case of a massless particle, which requires that one of the influence rates be zero.  That is, either $r_P = 0$ or $r_Q = 0$ so that $m = \sqrt{r_P r_Q} = 0$ and $E = |p|$.\footnote{It is important to note that the case where both rates are zero would result in not only zero mass but also zero energy and momentum.  Such a particle would not influence anything and would therefore be unobserved.}  In this case, the particle does not \emph{zitter}.  Instead, it influences only to the left or only to the right, which is described by observers as a massless particle traveling in one direction at the constant speed of light.  However, to obtain a rate of zero, the particle chain itself must project to a degenerate interval on one of the observer chains.

In relativistic quantum mechanics, this behavior of zig-zagging back-and-forth at the speed of light is predicted by the Dirac equation.  This predicted phenomenon was originally noted by Breit \cite{Breit:1928} and Schr{\"o}dinger \cite{Schrodinger1930:kraftefreie}, the latter of whom coined the term \emph{Zitterbewegung}, or `shuddering motion', which we will shorten to \emph{zitter}.  The phenomenon of \emph{zitter} was later emphasized by Huang \cite{Huang:1952zitterbewegung}, discussed by Feshbach and Villars \cite{Feshbach+Villars:1958}, described in a handful of texts \cite{Bjorken+Drell:1964}\cite{Feynman&Hibbs}\cite{Baym:1969}\cite{Merzbacher:1998}, and has been championed by Hestenes \cite{Hestenes:1990zitterbewegung}\cite{Hestenes:1993}\cite{Hestenes:2008electron}\cite{Hestenes:2010zitterbewegung}, Barut \cite{Barut+Bracken:1981:Zitter}\cite{Barut+Bracken:1981:magnetic-moment}\cite{Barut+Zanghi:1984} and others \cite{Gull+Lasenby+Doran:1993}\cite{Salesi+Recami:1994}\cite{Rodrigues+Vaz+Recami+Salesi:1998}\cite{Sidharth:2009} who have conceived of \emph{zitter} as a spiral motion and envisioned a connection to spin.  In the present context of 1+1 dimensions, \emph{zitter} manifests as a discrete zig-zagging motion, which, as we will discuss in Section \ref{sec:Feynman_Checkerboard}, is central to the Feynman checkerboard model of the electron \cite{Feynman&Hibbs}.
While \emph{zitter} has not yet been observed directly in electrons, there is not only indirect evidence that this is a real physical effect for electrons \cite{Gouanere+etal:2005:channeling}\cite{Catillon+etal:2008clock} but also evidence that this is a real effect for fermions in general \cite{Gerritsma+etal:2010quantum}\cite{Zawadzki+Rusin:2011}\cite{Leblanc+etal:2013:zitter}\cite{Qu+etal:2013:zitter}.

The phenomenon of \emph{zitter} should be of some concern since it challenges the concept of rest by implying that at the most fundamental level every particle is in a constant state of motion at the speed of light. That is, all particles, massive and massless, can \emph{only} go the speed of light!  All other speeds---including rest---are observed only on average.  As a result, \emph{zitter} goes as far as challenging the concept of a rest frame, which is central to the theory of general relativity.  This suggests that a theory such as general relativity can only hold on average in a coarse-grained picture and thus is likely to be inconsistent with quantum mechanics.

\subsection{Influence Sequences: Further Quantum Effects}
The proposal that a particle's position is defined, at least in part, by its influence on coordinated observers is a novel perspective, which may at first glance seem to run contrary to the usual idea that observers must make measurements, which affect the particle's behavior, to learn about a particle.  However, in this section we will show that the proposed model gives rise to several more quantum effects, including information isolation and the fact that measurements disturb particles.

The concept of position was derived as a consistent description of intervals between events in a coarse-grained setting.  We have seen in the previous section on \emph{Zitterbewegung}, that the description of intervals defined by influence events at the fundamental level leads to unexpected results that are consistent with some very poorly understood aspects of relativistic quantum mechanics.  Here we explore further consequences of the model.

\subsubsection{Compton Wavelength}
We begin by assuming that one knows the initial state of a free particle and consider the changes in the position (\ref{eq:directed_distance}) and the time (\ref{eq:duration}) used to describe the particle after it influences one of the observers.\footnote{By `position' and `time', we mean the directed distance and duration with respect to a defined origin.}  When the observer $\P$ is influenced, both the position and the time describing the particle change by $\frac{\Delta p}{2}$.  Similarly, influencing $\Q$ changes the position by $-\frac{\Delta q}{2}$, and the time by $\frac{\Delta q}{2}$.  In the rest frame, for any influence event, either $\Delta p = 1$ or $\Delta q = 1$.  Thus the time coordinate describing a particle advances in a discrete fashion by $\frac{1}{2}$, and its position can change by $\pm\frac{1}{2}$.  This means that the discrete nature of the act of influence imposes a fundamental unit of duration and distance beyond which one cannot measure.  One might postulate that this fundamental distance is related to the reduced Compton wavelength
\begin{equation}
\lambdabar = \frac{\hbar}{mc},
\end{equation}
which in natural units reduces to
\begin{equation}
\lambdabar = \frac{1}{m}.
\end{equation}
Given the fact that $m = 2$ (\ref{eq:mass=2}), we see that this is indeed the case. The reduced Compton wavelength is simply the smallest definable distance
\begin{equation}
\lambdabar = \frac{1}{2}.
\end{equation}

\subsubsection{Information Isolation and the Unorderability of Influence Sequences}
We now consider two coordinated observers, $\P$ and $\Q$, monitoring a free particle, represented by the chain of events $\Pi$.  The particle is assumed to be collinear to $\P\Q$, so that we are considering a 1+1 dimensional space induced by the two observers.
The observers record influence events and aim to make inferences about the sequence in which the particle has influenced the observers, which we refer to as an \emph{influence sequence}.  However, despite the fact that both $\P$ and $\Q$ have carefully recorded and quantified the influence events on their own chains, they are fundamentally unable to deduce the order in which the influence events occur on the particle chain $\Pi$.  This can be demonstrated by considering a simple example where more than one particle chain can give rise to the same influence events on the observer chains.

\begin{figure}[t]
  \begin{center}
  \includegraphics[height=0.4\textheight]{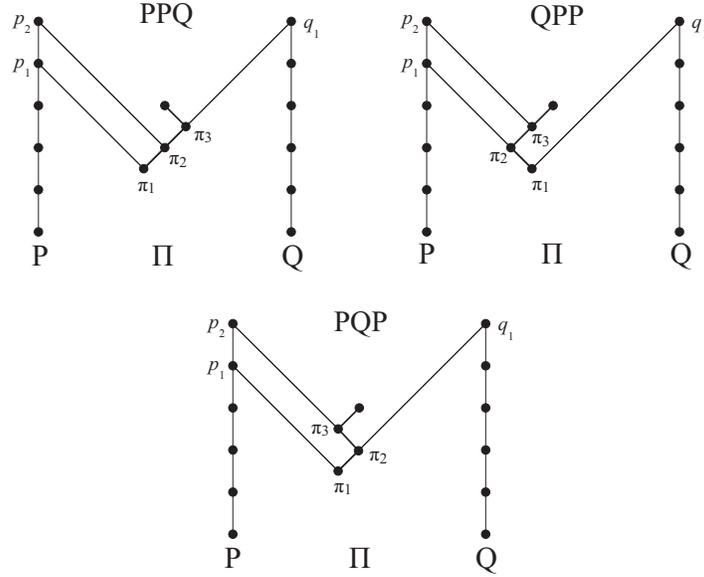}
  \end{center}
  \caption{Given just the influence events $p_1$, $p_2$, and $q_1$ associated with the observers $\P$ and $\Q$, there are three possible ways in which they could have resulted from the three acts of influence, $\pi_1$, $\pi_2$, and $\pi_3$ associated with the particle $\Pi$.  As a result, when making inferences about the behavior of the particle $\Pi$ as it evolves from its initial state $x_i$ (associated with $\pi_1$) to the final state (unlabeled), the observers must consider all possible orderings of influence instances.}
  \label{fig:fig-4}
\end{figure}

We consider a free particle that evolves from an initial state, represented by $X$, to a final state by influencing the observers $\P$ and $\Q$ at the events labeled $p_1$, $p_2$, and $q_1$ in Figure \ref{fig:fig-4}.  In this case, there are three possible orderings of the events on the free particle chain $\Pi$ that result in the same situation for the two observers.  In the first case, we have that the particle $\Pi$ influences $\P$ at $\pi_1$, $\P$ at $\pi_2$, and $\Q$ at $\pi_3$, which we write compactly as the influence sequence $\llseq{X}{P}{P}{Q}$.  The other two possible sequences, $\llseq{X}{P}{Q}{P}$ and $\llseq{X}{Q}{P}{P}$, are illustrated in Figure \ref{fig:fig-4}.  In general, one can show that the number of influence sequences to be considered is given by
\begin{equation}
S = \binom{N_p + N_q}{N_p} = \binom{N_p + N_q}{N_q} = \frac{(N_p + N_q)!}{N_p! N_q!},
\end{equation}
where $N_p$ and $N_q$ represent the number of times that the particle influences $\P$ and $\Q$.

Moreover, since each influence sequence of the particle corresponds to a different description of the particle by the observers, each distinct influence sequence corresponds to a distinct space-time path.  This is illustrated in Figure \ref{fig:fig-4} where the chain $\Pi$ is drawn specifically to depict the resulting space-time path in each case. As a result, the fact that the observers cannot determine the precise influence sequence of a free particle means that they cannot ascribe to that free particle a particular path through space-time.  Any inferences that the observers attempt to make about the free particle's behavior must take into account the set of possible influence sequences or, equivalently, the set of possible space-time paths.  This set of possible influence sequences constitute a set of \emph{interfering alternatives} \cite{Feynman&Hibbs}.

The fact that the observers are fundamentally unable to determine the precise influence sequence experienced by the free particle is an example of a key characteristic of quantum systems known as \emph{information isolation} \cite{Schumacher+Westmoreland:2010}.  Here, the information about the order in which the particle influences observers $\P$ and $\Q$ is inherently isolated from the observers.  There is no way for the observers to order these events.  Thus it is fundamentally impossible to describe the particle behavior precisely.  In this model, it is not that the free particle takes all possible paths through space-time.  Instead, because of information isolation, the observers must consider all possible paths to make optimal inferences about the particle.  The only question that remains is how does one consistently  take this information (or lack of information) into account to make such optimal inferences.

\subsubsection{Measurement}
\begin{figure}[t]
  \begin{center}
  \includegraphics[height=0.25\textheight]{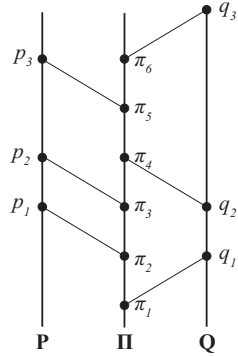}
  \end{center}
  \caption{This figure illustrates the process of measurement where by influencing the particle at the event $\pi_4$, the observer $\Q$, at least temporarily, disrupts information isolation and obtains information about the particle behavior.  In this case, the information comes in the form of a constraint, where it is learned that $q_2 \leq p_3$.  This limits the number of influence sequences that must be considered to make inferences about the particle $\Pi$.}
  \label{fig:fig-5}
\end{figure}

The fact that the influence sequence of a free particle is informationally isolated from the observers results in requiring that the observers consider all possible influence sequences when making inferences about the particle.  The situation changes if the observers themselves influence the particle.  Figure \ref{fig:fig-5} illustrates a particle that is influenced by the observer $\Q$ at event $\pi_4$ while it influences the observers.  The act of influencing the particle enables the observers to order some of the influence events along the particle chain $\Pi$.  For example, in this case it is known that event $q_2 \leq p_3$.  This constrains the possible set of influence sequences associated with the particle chain and at least partially destroys the information isolation by providing some information about the particle behavior.  However, in doing so, the observer necessarily influenced the particle.  We refer to this process as \emph{measurement}, as it results in providing information about the particle to the observers while affecting the particle behavior as one would expect in a quantum mechanical system.

\subsubsection{Quantifying the Free Particle State}
In this section we aim to describe how one can go about describing the free particle state, or influence sequence, for the purposes of making inferences.  Given that the free particle is fundamentally described by an influence sequence, we must focus on describing transitions from one influence event to another.  This results in four possible transitions:
\begin{align}
P \rightarrow P &\qquad Q \rightarrow P\\
P \rightarrow Q &\qquad Q \rightarrow Q.
\end{align}
We quantify each of these transitions with a complex number.  This is a big step and is justified, in part, by symmetries described in previous works \cite{GKS:PRA}\cite{GK:Symmetry}, which have been shown to be applicable to this problem \cite{Knuth:Info-Based:2014}.  Here we present an alternative perspective that is perhaps more intuitive.

We aim to quantify the behavior of the particle for the purposes of making inferences.  That is, we will assign a number or numbers to particular particle behaviors with the aim of consistently computing probabilities.  We have been considering the behavior of a free particle in terms of intervals defined by the transition from one influence event to another, which can be described by observers as a particle taking discrete steps in a space-time.  Alternatively, we could consider the behavior of the free particle in terms of rates, which can be described in terms of energy and momentum.  The two perspectives are Fourier duals to one another.  For this reason, the quantities that we use to describe the particle in these dual perspectives must also be Fourier duals to one another.  The simplest quantities that are assured to work in general are complex numbers.  However, there is a constraint that regardless of the perspective chosen, we must make consistent inferences.  That is, the probabilities computed using complex numbers describing influence sequence transitions must be equal to the probabilities computed from the Fourier transform of those complex numbers, which describe the influence sequence rates.  Parseval's theorem \cite{Arfken:1985} states that the only scalar derived from complex numbers that is invariant under the Fourier transform is the squared magnitude.  This is the Born Rule, simply conceived, as a consistency requirement arising from the fact that we must make consistent inferences when describing influence sequences in terms both of intervals and rates.

The fact that we are quantifying transitions from one influence event to another in a finite set of influence sequences implies that we must account for the fact that we will not, in general, possess complete information about any chosen `initial state'.  For this reason, we need to consider both the possibility that the particle has previously influenced $\P$ and the possibility that it has previously influenced $\Q$.  These two situations, which are analogous to arriving at the initial state from the left and right, respectively, are the two helicity states. To quantify them, we require two complex numbers:
\begin{equation}
\Phi =
\begin{pmatrix}
\phi_{\P} \\
\phi_{\Q}
\end{pmatrix}.
\end{equation}
The complex numbers $\phi_{\P}$ and $\phi_{\Q}$ are the quantum amplitudes for the particle to influence $\P$ and $\Q$, respectively.
The probability that the particle influences $\P$ is then given by
\begin{align}
\phi_{\P}^{*} \phi_{\P} &\equiv Prob(R) \nonumber \\
&= \frac{1+\langle \beta \rangle}{2} \nonumber \\
&= \frac{E+p}{2E}. \label{eq:(E+p)/2E}
\end{align}
Similarly, the probability that the particle influences $\Q$ is given by
\begin{equation}
\phi_{\Q}^{*} \phi_{\Q} = \frac{E-p}{2E}. \label{eq:(E-p)/2E}
\end{equation}
This allows us to write these numbers in a more familiar form
\begin{equation}
\phi_{\P} = \sqrt{\frac{E+p}{2E}} \, e^{i \theta_P} \qquad \phi_{\Q} = \sqrt{\frac{E-p}{2E}} \, e^{i \theta_Q}
\end{equation}
where $\theta_P$ and $\theta_Q$ are phase angles.

The result is that to make inferences about the influence patterns of a free particle, the observers must describe the particle with two complex numbers that take the form of a Pauli spinor.\footnote{Our initial studies of influenced particles indicate that one needs four complex numbers and that they appear to take the form of a Dirac spinor with the positive energy components representing the amplitudes for the particle to influence and the negative energy components representing the amplitudes for the particle to be influenced.}

In addition, it is important to note that since the state of a particle is defined in terms of the way in which the particle influences the observers, the fact that a particular space-time coordinate can be arrived at in only two ways, by influencing $\P$ or by influencing $\Q$, implies that at most two particles can occupy the same space-time coordinates and that these two particles must have opposite helicity states.  For example, consider the space-time position associated with the event $\pi_3$ in Figure \ref{fig:fig-4} in both the QPP and PQP cases.  In the QPP case, the particle arrives at the space-time position associated with $\pi_3$ by influencing $\P$ at the event $p_1$.  Whereas, in the PQP case, the particle arrives at the same space-time position by influencing $\Q$ at the event $q_1$.  Since that space-time position can only be reached by influencing $p_1$ or $q_1$, and since each event on the observer chain is defined by the influence of one and only one particle, this implies that no more than two particles can arrive at the same space-time position and that they must arrive with different helicity states.  Thus this model results in a 1+1-dimensional version of the \emph{Pauli Exclusion Principle}.

\subsection{The Feynman Checkerboard Model and the Dirac Equation}
\label{sec:Feynman_Checkerboard}
In 1965, Feynman and Hibbs \cite{Feynman&Hibbs} introduced the discrete one-dimensional checkerboard model of the electron where massive particles are conceived of as moving at the speed of light, but zig-zagging back-and-forth resulting in a subluminal average velocity.  This is typically illustrated in the standard space-time diagram as a path that zig-zags upwards along 45 degree angle light cones.

In the checkerboard model, each path segment is assigned a quantum amplitude so that the total amplitude of a given path (path amplitude) is found by taking the product of the amplitudes assigned to each segment comprising the path.  The amplitude assigned to the transition from some initial state at time $t_i$ and position $x_i$ to some final state at time $t_f$ and position $x_f$ is found by the discrete path integral summing the path amplitudes over all paths connecting $(t_i,x_i)$ to $(t_f,x_f)$.  It is posed as Problem 2-6 in Feynman and Hibbs \cite[p.35]{Feynman&Hibbs} to derive the Dirac equation by assigning an amplitude of $i \epsilon$ to path reversals and an amplitude of $\epsilon$ to straightaways.

Given the way in which space-time coordinates are assigned to influence events in this current theory, one can see that this model is isomorphic to the Feynman checkerboard (see Figure \ref{fig:fig-4} where the particle chain is illustrated as following a space time path).  Summing the amplitudes of space-time paths is equivalent to summing the amplitudes for all possible influence sequences.  We have previously shown that the amplitude assignments proposed by Feynman and Hibbs can be \emph{derived} using symmetries and consistency with probability theory \cite{Knuth:Info-Based:2014} in terms of transition matrices, or propagators, that evolve the spinor as the particle influences $\P$ or $\Q$:
\begin{equation}
P = \frac{1}{\sqrt{2}}
\begin{pmatrix}
1 & i \\
0 & 0
\end{pmatrix}
\qquad
Q = \frac{1}{\sqrt{2}}
\begin{pmatrix}
0 & 0 \\
i & 1
\end{pmatrix},
\end{equation}
so that
\begin{align}
P\Phi &= \frac{1}{\sqrt{2}}
\begin{pmatrix}
1 & i \\
0 & 0
\end{pmatrix}
\begin{pmatrix}
\phi_{\P} \\
\phi_{\Q}
\end{pmatrix}
\\
&= \frac{1}{\sqrt{2}}
\begin{pmatrix}
\phi_{\P} + i \phi_{\Q}\\
0
\end{pmatrix}
\end{align}
and
\begin{align}
Q\Phi &= \frac{1}{\sqrt{2}}
\begin{pmatrix}
0 & 0 \\
i & 1
\end{pmatrix}
\begin{pmatrix}
\phi_{\P} \\
\phi_{\Q}
\end{pmatrix}
\\
&= \frac{1}{\sqrt{2}}
\begin{pmatrix}
0\\
i \phi_{\P} + \phi_{\Q}.
\end{pmatrix}.
\end{align}
One can then readily derive the Dirac equation for the free particle as
\begin{align}
\partial_P \phi_{\P} &= i \phi_{\Q} \\
\partial_Q \phi_{\Q} &= i \phi_{\P},
\end{align}
which can be rewritten as
\begin{align}
(\partial_t + \partial_x) \phi_{\P} &= i \, m \, \phi_{\Q} \\
(\partial_t - \partial_x) \phi_{\Q} &= i \, m \, \phi_{\P},
\end{align}
where the mass $m=2$ as was found earlier (\ref{eq:mass=2}).

Observables can then be represented as matrix operators.  For example, the average velocity $\langle \beta \rangle$ of a particle can be found by defining the operator
\begin{equation}
\hat{\beta} =
\begin{pmatrix}
1 & 0 \\
0 & -1
\end{pmatrix}
\end{equation}
so that
\begin{align}
\langle \beta \rangle &= \Phi^\dagger \hat{\beta} \Phi \nonumber \\
&=
\begin{pmatrix}
{\phi_{\P}}^{*} & {\phi_{\Q}}^{*}
\end{pmatrix}
\begin{pmatrix}
1 & 0 \\
0 & -1
\end{pmatrix}
\begin{pmatrix}
\phi_{\P} \\
\phi_{\Q}
\end{pmatrix}
\nonumber \\
&=
\begin{pmatrix}
{\phi_{\P}}^{*} & {\phi_{\Q}}^{*}
\end{pmatrix}
\begin{pmatrix}
\phi_{\P} \\
-\phi_{\Q}
\end{pmatrix}
\nonumber \\
&= \phi_{\P}^{*} \phi_{\P} - \phi_{\Q}^{*} \phi_{\Q}.
\end{align}
This can also be written as $Prob(R) - Prob(L)$ as derived in (\ref{eq:beta_in_terms_of_probs}), or alternatively, by using (\ref{eq:(E+p)/2E}) and (\ref{eq:(E-p)/2E}), this can be written as $p/E$.

Here we see that the Feynman checkerboard model of the electron, which has been a curiosity of sorts attracting attention now-and-again throughout the years \cite{Gersch:1981feynman}\cite{Ord:1993dirac}\cite{Jacobson:1985}\cite{Kauffman:1996:DiscreteDirac}\cite{Gaveau+Schulman:2000}\cite{Ord:2002feynman}\cite{Kull:2002}\cite{Earle:DiracMaster2011}, is isomorphic to the simple model of an elementary particle that influences others in a direct and discrete fashion.

\section{Discussion}
The field of physics has been slowly constructed over the last four hundred years by identifying principles and relevant variables that aid in the optimal prediction of physical phenomena.  Since the discovery of the electron, physicists have struggled with the fact that several aspects of the mental models we use to conceive of an electron appear to be logically inconsistent despite the fact that the optimal predictive theory, known as quantum electrodynamics (QED), employs accepted relevant variables along with adopted principles to make the most accurate predictions of any physical theory in history. As a result, many foundational theorists work to develop sets of logically consistent principles by which quantum theory can be reconstructed.  While it is perhaps accepted that a successful theory will most likely need to revise or discard one or more commonly held beliefs, many foundational approaches attempt to retain as many familiar concepts and technical assumptions as possible so as to ensure success.  While this is a wise approach in some respects, it is not assured that it will result in what one might hope for in terms of a truly foundational theory.  This is because by assuming the relevance of a specific variable or adopting a specific principle or technical description, one is prevented from learning more about them.  For this reason, we have adopted a different foundational approach: build physics from the bottom up.

Rather than postulating laws and perceiving them as representing some kind of underlying natural order, we instead postulate the nature of the underlying order and derive the resulting laws.  This is accomplished by considering physics to represent a framework by which observers consistently quantify and make consistent optimal inferences about natural phenomena.  To do this successfully, it is important to look for clues. From our previous efforts \cite{Knuth:laws}\cite{Knuth:FQXI2015}\cite{Knuth&Skilling:2012}\cite{GKS:PRA}\cite{GK:Symmetry}\cite{Knuth+Bahreyni:JMP2014}\cite{Knuth:Info-Based:2014}\cite{Knuth:FQXI2013}\cite{Knuth:MaxEnt2014:motion}\cite{Walsh+Knuth:acceleration}, and those of others \cite{Cox:1946}\cite{Jaynes:Carnot}\cite{Pfanzagl:1968}\cite{Caticha:1998}, we have learned that symmetries inherent to a system constrain any attempt to consistently quantify that system.  That is, it is possible to begin with an underlying order and derive laws that are consistent with that order via the process of \emph{consistent quantification}.  This is the reason why mathematics is so successful at describing physics \cite{Knuth:FQXI2015}.  This is not idle philosophy.  We apply this critical observation to a simple model of an electron and use it to construct (reconstruct) a consistent physical theory from the bottom up.

We observe that the majority of the variables relevant to an electron which are often conceived of as representing properties of an electron, represent instead the relationship between the electron and an observer.  Based on this observation, we introduce the concept of influence as a simple means by which the electron-observer relationship is mediated.  It is postulated that an elementary particle, such as an electron, can influence another particle in a direct and discrete fashion.  Each instance of influence enables one to define two events, each associated with a different particle: the act of influencing and the act of being influenced.  Two additional postulates result in influence events being considered as a partially ordered set.  This allows us to describe a given particle as a totally ordered chain of events.

Consistent quantification of intervals defined by pairs of influence events results in the mathematics of space-time.  This implies that space and time need not be physical.  Instead, space and time are the uniquely consistent constructs by which one can describe events from the perspective of an embedded observer.  By considering events along a particle chain in terms of rates, we recover the concepts of mass, energy, and momentum, and note that these are necessarily Fourier duals of the invariant interval scalar, duration, and directed distance (position).  As a result the concept of complementarity emerges---not as complementarity among properties of particles, but rather as complementary descriptions of particles.

The fact that intervals and rates are Fourier duals suggests that the mathematical formalism of quantum mechanics might be derived by considering probabilities to be computed from quantities describing influence sequences.  The fact that one's inferences should be invariant with respect to one's description of a system in terms of intervals or rates, suggests that such systems should be quantified by complex numbers and that the Born Rule is simply an example of Parseval's theorem applied to influence sequences.

In addition to complementarity, several other concepts central to quantum mechanics emerge naturally from the model: information isolation, the Compton wavelength, and the Pauli exclusion principle.  Optimal inferences about the behavior of a free particle result in the Dirac equation with Pauli spinors used to quantify the particle behavior in 1+1-dimensions.  The proposed model exhibits \emph{Zitterbewegung}, which is a poorly understood relativistic quantum effect predicted by the Dirac equation and intimately related to mass, spin, and velocity.
It is presently thought that the Higgs field gives rise to mass. However, this does not explain the intimate relationship between mass and spin, which has been investigated by Hestenes and others.
From the perspective of the proposed theory, \emph{Zitterbewegung}, mass, and (at least) helicity (a 1+1-dimensional analog of spin) arise from the fact that particles influence one another.  This leads one to wonder if the Higgs field simply represents this network of influence instances.

Despite the fact that a surprising amount of physics can be derived from this simple model of one particle influencing another, it would be naive to assume that the ideas presented here comprise anything resembling the final word on the matter.  It is hoped that this work has demonstrated that a simple and understandable picture of particles, such as the electron, as well as a broad and coherent foundational theory of physics are indeed feasible and should be actively sought after.

\section*{Acknowledgements}
I would like to thank Newshaw Bahreyni, Seth Chaiken, Ariel Caticha, Keith Earle, David Hestenes, Oleg Lunin, John Skilling, and James Lyons Walsh for numerous insightful discussions.  I also want to specifically thank James Lyons Walsh for his careful proofreading of this manuscript and his invaluable comments.

\bibliographystyle{amsplain}
\bibliography{C:/Users/KK952431/kevin/files/papers/bibliography/knuth}

\end{document}